\documentclass[11pt]{article}
\usepackage[margin=1in]{geometry}
\usepackage{amsmath,amssymb,graphicx,booktabs,array}
\usepackage[numbers,sort&compress]{natbib}
\usepackage{setspace}
\usepackage[hidelinks]{hyperref}

\newcommand{\ISF}{\mathrm{ISF}}
\newcommand{\CF}{\mathrm{CF}}
\newcommand{\HGP}{\mathrm{HGP}}

\title{Sliding-scale Insulin Does Not Control Steroid-induced Hyperglycemia in a Non-diabetic Patient, No Matter How You Tune It}
\author{Nir Regev$^{1,*}$ and Guy Regev$^{1}$}
\date{}

\begin{document}
\maketitle
\noindent $^{1}$Regev Labs, LLC, Temecula, CA, USA.\\
$^{*}$Corresponding author: nir@alephzero.ai

\section*{Abstract}
Sliding-scale insulin is the default hospital response to high blood glucose. Its correction factor was designed for patients who make no insulin, yet it is routinely ordered for patients whose pancreas still works, including those on high-dose steroids. We treat the sliding scale as what it is, a proportional controller sampled every four hours, acting on a patient who already contains a working glucose controller of her own. We test it in a non-diabetic adolescent on high-dose corticosteroids and parenteral nutrition, and in 4,995 virtual patients. It fails for three structural reasons, and no choice of correction factor fixes any of them. First, the patient's own loop settles in 11.5 minutes while injected insulin takes 64 minutes to peak, so most of each dose simply replaces insulin she would have secreted; in simulation the ordered scale moved her mean glucose by less than 1.5~mg/dL. Second, one unit lowers glucose twice as much at 110~mg/dL as at 190, so any correction factor strong enough to work when glucose is high overshoots when it is low; in 99.5\% of virtual patients no constant factor does both. Third, four-hour sampling is far below the Nyquist rate, so successive readings are uncorrelated and each dose responds to a value that does not persist. Even the argument that correction doses rest the beta cells, taken on its own terms, favors continuous delivery. We propose replacing the scale with a continuous insulin infusion co-delivered with the dextrose and adjusted once a day on mean glucose, the one quantity four-hour monitoring can measure.

\section*{Introduction}
Correction insulin given on a sliding scale remains the most common inpatient response to hyperglycemia. Current standards discourage its sole use\cite{ADA2026}, but the supporting evidence comes almost entirely from adults with type 2 diabetes\cite{Umpierrez2007}, and the arithmetic behind the correction factor was developed for basal-bolus dosing in patients who depend on exogenous insulin\cite{Davidson2008}.

A large inpatient population sits outside that evidence: patients with working beta cells made hyperglycemic by glucocorticoids\cite{Radhakutty2018}, parenteral nutrition\cite{Schonenberger2022}, or both. In these patients the injected dose does not act on a passive system. It acts on a plant that already contains a feedback controller, the pancreas, which responds to any fall in glucose by reducing its own secretion. The effect of this internal loop on exogenous insulin has been studied in closed-loop systems that sample continuously\cite{Steil2011,Boughton2019}, but not for the four-hourly proportional rule that is actually ordered on most wards.

Here we treat the sliding scale as what it is in control terms, a sampled proportional controller with a deadband, and ask whether any choice of its gain can work when the patient's own loop is intact. We derive the sensitivity of glucose to exogenous insulin, evaluate the resulting per-unit potency in a virtual population drawn from published physiological ranges, and calibrate the model to a non-diabetic adolescent after allogeneic stem cell transplantation, treated with high-dose methylprednisolone and total parenteral nutrition.

We find that the failure is structural rather than a matter of tuning. The endogenous loop rejects most of each dose, so the ordered rule has almost no authority over mean glucose. Per-unit potency falls as glucose rises, so a correction factor large enough to be safe at low glucose is ineffective at high glucose; how large this mismatch is depends mainly on where the beta cell dose-response curve is centered. Four-hour sampling is far below the Nyquist rate of these dynamics, so successive readings are uncorrelated and each dose responds to a value that does not persist. The argument that correction doses rest the beta cells, taken on its own terms, favors continuous delivery. We propose a continuous insulin infusion co-delivered with the dextrose and updated once daily on the previous day's mean glucose.

\section*{Results}

\subsection*{The ordered rule is a sampled proportional controller}
Doses were computed by the hospital's electronic inpatient insulin calculator. No public technical description of that calculator exists, so the structure below is reconstructed from the charted doses and from standard basal-bolus arithmetic\cite{Davidson2008}. The nurse enters the point-of-care glucose and a carbohydrate figure, which for this patient is always zero because she takes nothing by mouth. With the carbohydrate and ketone terms inactive the calculator reduces to
\begin{equation}
u_c = \mathrm{quant}\!\left[\frac{(\tilde G - T)^+}{\CF}\right], \qquad \tilde G = G(t_c) + \varepsilon, \qquad t_{c+1} - t_c \approx 4~\mathrm{h},
\label{eq:rule}
\end{equation}
where $u_c$ is the dose in units, $\tilde G$ the measured glucose including meter error $\varepsilon$, $T$ the subtracted target, $\CF$ the correction factor in mg/dL per unit, and $\mathrm{quant}[\cdot]$ floors to half units with a half-unit minimum. Three administrations pin the parameterization. A reading of 177~mg/dL at 08:24 on day +50 after transplant produced 1.0 unit; 188 at 08:37 on day +51 produced 1.5 units; and 173 at 20:56 on day +49 produced nothing. With the ordered $\CF = 40$, the two daytime doses place the target near the midpoint of the charted 100 to 150~mg/dL band ($T = 125$), and the undosed overnight reading requires an overnight target above 153, consistent with the top of the 125 to 175 band ($T = 175$). The half-unit floor means nothing is given below 145~mg/dL by day or 195 overnight, which is why the rule is described at the bedside as a threshold of 150. The divisor had previously been 60 and then 45.

In control terms Eq.~(\ref{eq:rule}) is a proportional controller with a deadband, sampled every four hours, whose gain is $1/\CF$. Whether any value of $\CF$ can work depends on how much one unit actually moves glucose, which we derive next.

\subsection*{The endogenous loop rejects injected insulin}
The model (Methods, Eqs.~\ref{eq:model} and \ref{eq:secretion}) is a single glucose compartment with dextrose appearance, hepatic output suppressed by insulin, insulin-independent uptake, glucose effectiveness, renal loss above a threshold, and insulin-mediated disposal. Endogenous secretion is a saturating sigmoid in glucose with a 0.4~h lag. Linearizing about an operating point and eliminating the pancreatic response gives the steady-state change in glucose per unit per hour of sustained exogenous insulin,
\begin{equation}
\frac{dG}{du} = -\frac{A}{\underbrace{A\,S'(G)}_{\text{pancreas}} + \underbrace{U_{ni}'(G) + S_G}_{\text{insulin-independent}} + \underbrace{R'(G)}_{\text{kidney}}}, \qquad A \equiv \ISF + \left|\frac{\partial \HGP}{\partial I}\right|,
\label{eq:sens}
\end{equation}
where $A$ is the total marginal effect of insulin on net glucose flux, through peripheral uptake ($\ISF$) and through suppression of hepatic output; $S'(G)$ is the slope of the beta cell dose-response curve; $U_{ni}'$ and $S_G$ are the slopes of saturable uptake and of glucose effectiveness; and $R'$ is the renal slope, zero below the threshold. The denominator is the total restoring rate $\lambda$, whose reciprocal is the closed-loop time constant. The form is the standard disturbance-rejection result: from the patient's point of view the injected insulin is a disturbance, and a plant that regulates itself at rate $\lambda$ rejects it in proportion.

For a single subcutaneous bolus the closed-loop time constant is much shorter than the time to peak insulin action, so the loop tracks the bolus quasi-statically and the peak glucose fall from one unit is approximately
\begin{equation}
k(G) \simeq \left|\frac{dG}{du}\right| h(t^*), \qquad h(t) = \frac{e^{-t/\tau_{\mathrm{off}}} - e^{-t/\tau_{\mathrm{on}}}}{\tau_{\mathrm{off}} - \tau_{\mathrm{on}}},
\label{eq:potency}
\end{equation}
with $h$ the unit-area impulse response of subcutaneous insulin action and $t^* = \ln(\tau_{\mathrm{off}}/\tau_{\mathrm{on}})/(1/\tau_{\mathrm{on}} - 1/\tau_{\mathrm{off}}) = 1.07$~h (64 minutes), $h(t^*) = 0.279~\mathrm{h}^{-1}$. $k(G)$ is precisely the quantity the ordered correction factor claims to be. Equation~(\ref{eq:potency}) ignores the overshoot of the closed loop and underestimates the simulated peak by about 20\% at the operating point (6.4 against 8.0~mg/dL per unit); all values of $k$ reported below are computed by simulating the full nonlinear model.

Calibrated to the index patient at 6~g/kg/day of dextrose (289~g/day, a glucose appearance of 165.5~mg/dL/h), the model with no exogenous insulin settles at 139~mg/dL, with hepatic output $+25.9$, brain and red cell uptake $-45.9$, renal loss 0 and insulin-mediated disposal $-145.6$~mg/dL/h, and endogenous secretion of 1.53~U/h, or 37 units a day. Normal endogenous secretion measured by C-peptide deconvolution is $63 \pm 15$~U/day in healthy adults on a high-carbohydrate diet\cite{Kruszynska1987,Polonsky1988}, roughly $0.9 \pm 0.2$~U/kg/day or $41 \pm 10$~U/day at her weight. She secretes a normal amount against an abnormal glucose input with abnormal resistance, which is why the glucose sits high rather than the insulin sitting low. The 15 checks taken at this dextrose averaged 145~mg/dL.

At this operating point $A = 120$~mg/dL/U, $S' = 0.0307$~U/h per mg/dL, $U_{ni}' + S_G = 1.54~\mathrm{h}^{-1}$ and $R' = 0$, giving $\lambda = 5.22~\mathrm{h}^{-1}$ ($1/\lambda = 11.5$~minutes) and $dG/du = -23.0$~mg/dL per U/h. The pancreas supplies 70\% of the restoring rate. Most of an injected dose is therefore substituted rather than added: one unit per hour lowers glucose by 23~mg/dL, that fall reduces endogenous secretion by $S' \times 23 = 0.70$~U/h, and the total insulin acting rises by only 0.30~U/h. The residual glucose fall is not a leak in the compensation; it is the signal that causes it.

In stochastic simulation of the ordered rule over nine days (Methods), the rule delivered 0.25~U/day, fired on 8\% of checks and moved mean glucose by 0.2~mg/dL with a deterministic plant; with a slow sensitivity disturbance fitted to the observed spread of the checks it delivered 0.95~U/day, fired on 22\% of checks and moved the mean by 0.7~mg/dL (Table~\ref{tab:sim}). At 8~g/kg/day of dextrose the authority rose only to 1.1 to 1.4~mg/dL.

\begin{table}[h]
\centering
\caption{\textbf{Sampled-loop simulations.} 300 runs per row, nine days after a one-day burn-in (thirteen for the interruption rows); checks near 00, 04, 08, 12, 16 and 20~h with uniform $\pm 0.5$~h jitter and meter error of SD 7~mg/dL. ``Deterministic'' has no sensitivity disturbance; ``fitted'' adds a slow log-sensitivity disturbance matched to the within-day spread of the observed checks. Minima are medians over runs of the true trajectory minimum.}
\label{tab:sim}
\begin{tabular}{lrr}
\toprule
& deterministic & fitted\\
\midrule
\multicolumn{3}{l}{\emph{Ordered rule, 6 g/kg/day}}\\
\quad insulin delivered (U/day) & 0.25 & 0.95\\
\quad checks triggering a dose & 8\% & 22\%\\
\quad change in mean glucose (mg/dL) & 0.2 & 0.7\\
\quad same, 8 g/kg/day & 1.1 & 1.4\\
\multicolumn{3}{l}{\emph{Correction factor sweep: SD at checks / nine-day minimum (mg/dL)}}\\
\quad $\CF = 40$ & 8.1 / 124 & 25.7 / 61\\
\quad $\CF = 11.6$ & 8.2 / 122 & 25.8 / 57\\
\quad $\CF = 5.8$ & 8.5 / 104 & 26.4 / 42\\
\quad $\CF = 2.9$ & 9.8 / 63 & 28.7 / $<40$\\
\multicolumn{3}{l}{\emph{8 g/kg/day, unbridged interruptions: last-3-day mean / runs below 70}}\\
\quad no insulin & 155 / 0\% & 155 / 96\%\\
\quad scale, $\CF = 40$ & 154 / 0\% & 154 / 98\%\\
\quad infusion, separate line & 144 / 28\% & 145 / 100\%\\
\quad infusion, co-delivered & 144 / 0\% & 145 / 100\%\\
\bottomrule
\end{tabular}
\end{table}

\paragraph{The record shows the same thing.} On the night of day +49 a reading of 173~mg/dL at 20:56 fell under the overnight threshold and was not corrected. With no insulin given, the next readings were 154 at 01:04, 119 at 04:15 and 116 at 04:18, before 177 at 08:24 triggered one unit. Glucose fell 57~mg/dL in 7.4 hours unaided and stopped at 116 rather than overshooting. At the modeled potency near 160~mg/dL (6.9~mg/dL per unit) that is what roughly eight injected units would have done.

\paragraph{The argument requires that the pancreas is not near saturation,} since $S'$ collapses at the ceiling of the sigmoid. Secretory capacity and $\ISF$ enter the model only as a product and are not separately identifiable from glucose levels, so we test saturation from the response to the carbohydrate load. If the pancreas were saturated, the only restoring term would be the insulin-independent one at $1.54~\mathrm{h}^{-1}$, and glucose would move by 17.9~mg/dL per g/kg/day of dextrose; with the pancreas responding the linearized model gives 5.3 and the full model 7.2. Dextrose fell by 2~g/kg/day over the recorded week, so the saturated hypothesis predicts a fall of 36~mg/dL and the regulated one 11 to 14. The per-day inversion reported below attributes 19~mg/dL of the observed change to the dextrose, nearer the regulated prediction. We do not claim this discriminates sharply; the stronger argument is that saturation would require her maximum secretory capacity to equal her present output of 37 units a day, within the normal range of $41 \pm 10$, in a 17-year-old whose pre-transplant glycated hemoglobin was 4.9\%.

\subsection*{Per-unit potency varies twofold across the treated range}
\begin{figure}[t]
\centering
\includegraphics[width=\textwidth]{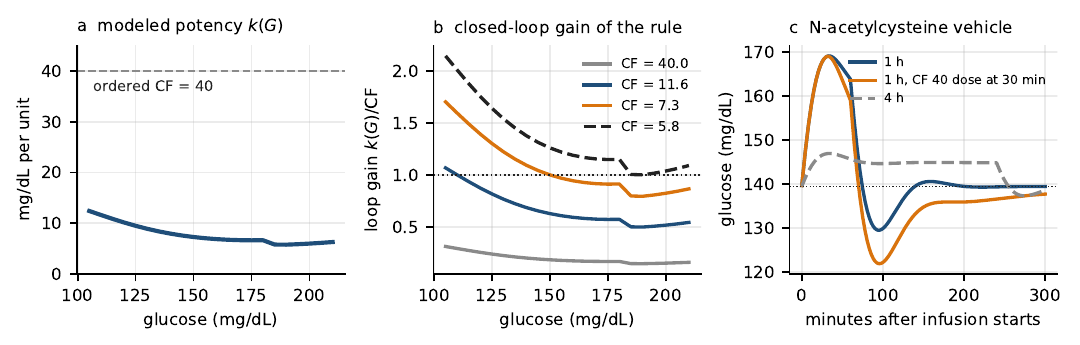}
\caption{\textbf{Per-unit potency, closed-loop gain and the dextrose-vehicle pulse in the index patient.} \textbf{a} Modeled peak glucose fall per unit, $k(G)$, from the full nonlinear model, with the ordered correction factor of 40~mg/dL per unit (dashed). \textbf{b} Closed-loop gain $k(G)/\CF$ for four correction factors: the ordered 40, and the values that give unit gain at 110, 150 and 190~mg/dL. The dotted line is unity, above which each correction overshoots its own target. \textbf{c} Glucose after the 8.35~g of dextrose that carries intravenous N-acetylcysteine, infused over one hour or over four hours, and over one hour with the correction dose the ordered rule gives if a check lands 30 minutes into the infusion.}
\label{fig:potency}
\end{figure}

Evaluating $k(G)$ by simulation at each operating point, with dextrose adjusted to hold it, gives the per-unit potency and hence the closed-loop gain $k(G)/\CF$ of the correction rule (Table~\ref{tab:potency}, Fig.~\ref{fig:potency}). A gain above one means each correction overshoots its own target.

\begin{table}[h]
\centering
\caption{\textbf{Modeled per-unit potency and closed-loop gain of the correction rule.} $k(G)$ is the peak glucose fall from one subcutaneous unit given at steady state $G$, from the full nonlinear model; the renal-off column removes the renal term. Gains are $k(G)/\CF$.}
\label{tab:potency}
\begin{tabular}{rrrcccc}
\toprule
$G$ (mg/dL) & $k(G)$ & renal off & $\CF = 40$ & $\CF = 11.6$ & $\CF = 7.3$ & $\CF = 5.8$\\
\midrule
110 & 11.6 & 11.6 & 0.29 & 1.00 & 1.59 & 2.01\\
130 & 8.9 & 8.9 & 0.22 & 0.77 & 1.22 & 1.54\\
150 & 7.3 & 7.3 & 0.18 & 0.63 & 1.00 & 1.26\\
170 & 6.7 & 6.7 & 0.17 & 0.58 & 0.91 & 1.15\\
190 & 5.8 & 6.8 & 0.14 & 0.50 & 0.79 & 1.00\\
210 & 6.3 & 7.5 & 0.16 & 0.54 & 0.86 & 1.09\\
\bottomrule
\end{tabular}
\end{table}

Potency rises by a factor of 2.0 between 190 and 110~mg/dL. Two mechanisms produce it. The first is the shape of the beta cell dose-response: $S'$ is largest at the secretion midpoint $G_{50}$ and falls away on both sides, so the pancreas opposes an injected dose most strongly near $G_{50}$. The second is the kidney: above the renal threshold of 180~mg/dL the renal slope joins the restoring rate, which is why $k$ drops from 6.7 at 170 to 5.8 at 190 instead of rising. Without the renal term $k(190) = 6.8$ and the ratio is 1.7. The consequence is a conflict between two requirements:
\begin{itemize}
\item to correct a reading of 190~mg/dL, the correction factor must be near $k(190) = 5.8$~mg/dL per unit (6.8 without the renal term);
\item to keep the closed-loop gain at or below one everywhere in range, it must exceed $k(110) = 11.6$~mg/dL per unit.
\end{itemize}
These are incompatible by a factor of 2.0, or 1.7 with the renal term removed. No constant correction factor satisfies both. The ordered $\CF = 40$ avoids the conflict by being 3.4 to 6.9 times larger than the modeled potency at every glucose, which is why successive tightening from 60 to 45 to 40 produced no visible change. The scale is safe because it does nothing, and the two facts have the same cause.

\paragraph{What a gain above one does and does not mean.} It is an overshoot criterion, not a stability criterion. The loop returns to its operating point within minutes, so by the next four-hour check the perturbation from any single dose is gone. The sampled sequence diverges only when the residual action of previous doses is amplified, which for these insulin kinetics requires a gain above 3.3, a correction factor below 3.5. What changes before that point is the depth of each transient, which the sampler does not record. With a deterministic plant the spread of the checks barely changes as the correction factor falls from 40 to 5.8 (standard deviation 8.1 to 8.5~mg/dL), while the true nine-day minimum falls from 124 to 104~mg/dL, and to 63 at 2.9 (Table~\ref{tab:sim}).

\paragraph{Why a correction factor tuned for effectiveness is nevertheless unsafe here.} On a representative reading it is not: at $\CF = 5.8$ a sustained 190~mg/dL calls for 11 units, about seven hours of her entire endogenous output given at once, and the modeled nadir is 117. The difficulty is that four-hour sampling cannot establish that a reading is representative. A check landing on the peak of the morning dextrose-vehicle pulse (below) reads 186 while the underlying glucose is about to fall, and the same rule then gives the doses in Table~\ref{tab:pulse}. A correction factor tuned to be effective is safe on a reading that represents a sustained level and dangerous on one that does not, and the monitoring cannot tell the two apart.

\begin{table}[h]
\centering
\caption{\textbf{A check landing on a vehicle pulse.} The reading is taken 30 minutes into the one-hour morning infusion of N-acetylcysteine and levofloxacin in 5\% dextrose; nadirs are modeled. The last column gives the nadir if the same dose were given at the sustained operating point of 139~mg/dL. Nadirs below 40~mg/dL are reported as $<40$ because the model is not valid there.}
\label{tab:pulse}
\begin{tabular}{rrrr}
\toprule
$\CF$ & dose at a reading of 186 (U) & modeled nadir & nadir from a sustained 139\\
\midrule
40 & 1.5 & 112 & 127\\
11.6 & 5.0 & 83 & 96\\
5.8 & 10.5 & $<40$ & $<40$\\
4.0 & 15.0 & $<40$ & $<40$\\
\bottomrule
\end{tabular}
\end{table}

\subsection*{The potency mismatch across a virtual population}
\begin{figure}[t]
\centering
\includegraphics[width=\textwidth]{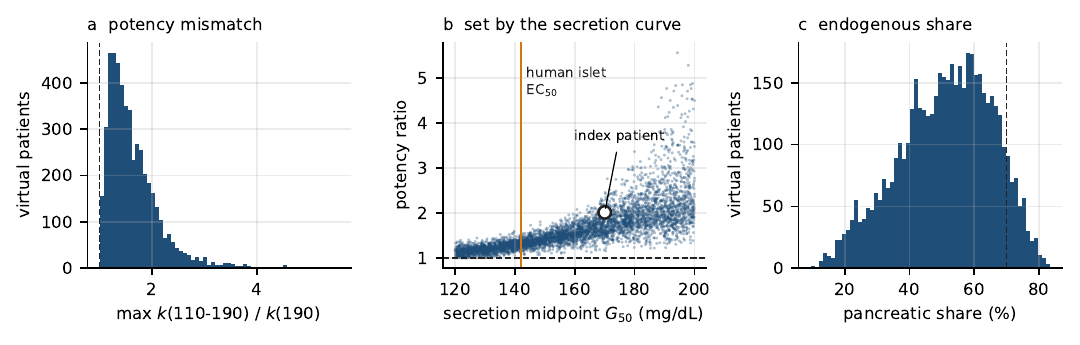}
\caption{\textbf{Virtual population of 4,995 patients with intact secretion.} \textbf{a} Potency ratio, the largest modeled $k$ between 110 and 190~mg/dL divided by $k(190)$; above one (dashed), no constant correction factor is both effective at 190 and free of overshoot below it. \textbf{b} Potency ratio against the secretion midpoint $G_{50}$. The vertical line marks the half-maximal glucose of perifused human islets\cite{Buchwald2019}; the open circle is the index patient. \textbf{c} Pancreatic share of the restoring rate, $A S'/\lambda$, which is also the fraction of each exogenous dose that is substituted; the dashed line is the index patient.}
\label{fig:vpop}
\end{figure}

The index patient is one parameter set. To ask whether the findings hold in patients with intact secretion generally, we sampled 5,000 virtual patients from published ranges for body weight, glucose distribution volume, glucose effectiveness, insulin-independent uptake, hepatic output, the position and width of the secretion curve, secretion lag, subcutaneous insulin kinetics, the glucocorticoid reduction in insulin action, renal function and daily secretion, and calibrated each to its own operating point between 130 and 200~mg/dL (Methods). Five were excluded for an implausible implied dextrose load, leaving 4,995 (Fig.~\ref{fig:vpop}).

The first finding generalizes without exception. The closed-loop time constant was 17.6 minutes (median; 5th to 95th percentile 11.0 to 26.2), the pancreas supplied a median 53\% (25 to 73\%) of the restoring rate, and the same fraction of each exogenous dose is substituted. The ordered correction factor of 40 exceeded $k(190)$ in every virtual patient, by a median factor of 8.5, so the ordered rule is inert across the population. The index patient sits toward the fast end: her 70\% pancreatic share is at the 91st percentile and her 11.5-minute time constant at the 7th.

The second finding holds in direction and varies in size. The potency ratio, the largest $k$ between 110 and 190~mg/dL divided by $k(190)$, had a median of 1.51 (1.11 to 2.62). It exceeded one, so that no constant correction factor is both effective at 190 and free of overshoot below it, in 99.5\% of virtual patients (91\% with the renal term removed); it exceeded 1.25 in 79\% (59\%) and 1.5 in 51\% (36\%). Partial rank correlation identified the secretion midpoint $G_{50}$ as the dominant determinant ($\rho = 0.92$; next largest, the operating point, $-0.32$). The median ratio was 1.19 for $G_{50}$ between 120 and 140~mg/dL, 1.40 for 140 to 160, 1.71 for 160 to 180 and 2.06 for 180 to 200.

This makes $G_{50}$ the parameter the second finding rests on. Perifused human islets reach half-maximal secretion at 7.9~mM, about 142~mg/dL\cite{Buchwald2019}, below the 170~mg/dL used for the index patient. In the 933 virtual patients with $G_{50}$ between 135 and 150~mg/dL the median ratio was 1.31 (1.15 to 1.51); it exceeded 1.25 in 73\%, but in only 10\% once the renal term was removed. At an islet-like secretion midpoint, therefore, the gap between the effective and the safe correction factor is about 30\%, and about half of it comes from the kidney above 180~mg/dL (median ratio 1.15 with the renal term removed). The twofold ratio in the index patient is at the 83rd percentile of the population and should be read as her value, not a general one.

\subsection*{Four-hour sampling aliases a twelve-hour process}
\begin{figure}[t]
\centering
\includegraphics[width=\textwidth]{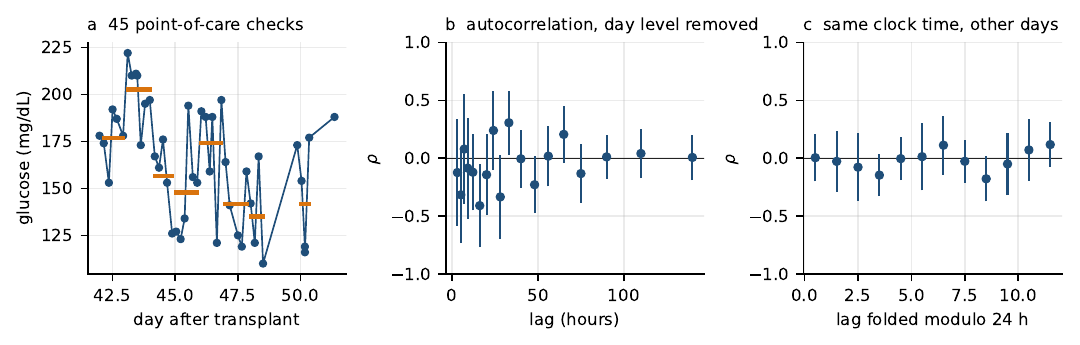}
\caption{\textbf{The point-of-care record and its correlation structure.} \textbf{a} The 45 checks from day +41 to day +51 after transplant, with each day's mean. \textbf{b} Autocorrelation of the within-day residual against true elapsed lag, with $\pm 2$ standard errors. \textbf{c} The same residual with the lag folded modulo 24 hours.}
\label{fig:record}
\end{figure}

Taken in aggregate, the 45 readings carry no memory. After removing each day's mean, the lag-one correlation at the nominal four-hour spacing is $-0.14$, against a 95\% band of $\pm 0.32$. A slotted correlogram over the true irregular spacings gives 18 bins covering lags up to 160 hours, of which two marginally exceed their own two-standard-error band, at 16 and 33 hours and with opposite signs, as expected from multiple comparisons. Folding the lag modulo 24 hours gives twelve bins, none of which departs from zero (Fig.~\ref{fig:record}).

The 24-hour fold is the wrong period. Methylprednisolone is given twice daily, so the driving rhythm is 12-hourly. Folding at 12 hours on the 15 checks taken at the current dextrose gives $\rho = +0.62$ at zero clock separation and $-0.72$ at half a period. Over all 45 checks, across which the dextrose changed twice, the same fold shows no structure ($+0.06$ and $-0.10$). A slow drift decays monotonically and never goes negative; only a periodic process changes sign at half a period, so the negative value distinguishes a rhythm from persistence. The half-period estimate rests on 10 pairs and is exploratory. The process is cyclostationary, and the ordinary autocorrelation, which averages over the cycle, returns zero by construction.

Three consequences follow. The difference between two consecutive readings is not a dose response: at this sampling interval it is dominated by where in the 12-hour cycle each reading fell. An excursion lasting $D$ minutes is recorded with probability of about $D/240$, so a 30-minute dip appears about one time in eight. And in simulation with the fitted disturbance, four-hour sampling recovers the mean to within 3~mg/dL and the standard deviation to within 2, but overestimates the true minimum by a median of 18~mg/dL.

\subsection*{Dextrose in medication diluents}
Seven percent of the carbohydrate load is not in the nutrition order. Intravenous N-acetylcysteine is diluted in 167~mL of 5\% dextrose and infused over one hour at 06:00 and 18:00, delivering 8.35~g of glucose in that hour against a nutrition baseline of 12.0~g/h, a 69\% step twice a day. Levofloxacin is a 100~mL premix in 5\% dextrose, also infused over one hour, once daily at 06:00, so the morning hour carries 13.35~g, more than the parenteral nutrition itself. A third medication, ganciclovir, contributed a further 6~g per dose until its diluent was changed to saline on day +48.

Spread evenly, these 21.7~g a day would be worth 2.4~mg/dL on the mean. Delivered as one-hour boluses, the N-acetylcysteine alone produces a modeled excursion of 30~mg/dL peaking about 34 minutes after the infusion starts, followed by a rebound to 10~mg/dL below baseline about half an hour after it ends, as the insulin secreted during the pulse outlasts the sugar that provoked it (Fig.~\ref{fig:potency}c). With levofloxacin added the morning excursion is 47~mg/dL and the rebound 16. A check landing on the morning peak triggers, at the ordered correction factor, a dose whose action arrives on the trough and takes the nadir from 124 to 112~mg/dL; at correction factors tuned to be effective it is far lower (Table~\ref{tab:pulse}). Lengthening the same infusion to two hours cuts the N-acetylcysteine excursion to 15~mg/dL and to four hours cuts it to 8, at no cost and with no change of diluent.

None of this appears in the glucose record. A one-hour excursion is far shorter than the four-hour sampling interval, so the checks sample it well below its Nyquist rate and record it only by chance. A single fingerstick about 40 minutes into the morning infusion would measure an excursion the monitoring has not recorded, and would confirm or refute the estimate above at the cost of one stick.

\subsection*{Correction doses give little beta cell relief}
A separate argument for the correction scale is that it rests the beta cells: sustained hyperglycemia degrades insulin sensitivity and secretion, so exogenous insulin reduces the secretory burden whether or not it moves mean glucose. The evidence for glucotoxicity and beta cell rest concerns marked sustained hyperglycemia and substantial insulin exposure, not about a unit a day, and no clinical benefit at these doses has been shown. We nevertheless answer it on its own terms, by running the model with and without each administered dose and integrating the difference in endogenous secretion over the following eight hours,
\begin{equation}
\Delta_{\mathrm{spared}} = \int_{t_c}^{t_c + 8\,\mathrm{h}} \left[I_e^{(\text{no dose})}(t) - I_e^{(\text{dose})}(t)\right] dt,
\end{equation}
the number of units of her own insulin she did not have to make (Table~\ref{tab:spare}).

\begin{table}[h]
\centering
\caption{\textbf{Endogenous insulin spared by the two administered doses.} Eight-hour totals are taken at glucose of 177 to 188~mg/dL, above the mean, so the implied secretion rate is higher than the daily average.}
\label{tab:spare}
\begin{tabular}{lrr}
\toprule
& 1.0 U at 177 & 1.5 U at 188\\
\midrule
Secretion over the next 8 h, no dose (U) & 22.87 & 26.03\\
Secretion with the dose (U) & 22.20 & 25.16\\
Endogenous insulin spared (U) & 0.68 & 0.88\\
\quad as a fraction of the dose given & 68\% & 58\%\\
\quad as a fraction of the 8 h workload & 3.0\% & 3.4\%\\
Secretion rate at peak exogenous action (U/h) & 2.67 & 3.01\\
Exogenous delivery rate at that moment (U/h) & 0.28 & 0.42\\
Peak instantaneous share carried by the dose & 9\% & 12\%\\
\bottomrule
\end{tabular}
\end{table}

Between 58 and 68\% of each dose replaces insulin she would otherwise have secreted; as beta cell sparing that substitution is efficient. At the peak of insulin action, about 64 minutes after injection, the syringe carries 9 to 12\% of the insulin reaching tissue, the strongest form in which the argument can be made. Globally the relief is small because the window is short: sparing 0.7 to 0.9 of the 23 to 26 units secreted over eight hours is 3.0 to 3.4\% of the work, and between doses the relief is zero. If protecting the beta cells is the objective, the relevant comparison is not a dose against none but a spike against a continuous term. At the current dextrose a continuous infusion of 7~U/day (0.29~U/h) carries 18\% of the insulin reaching tissue at all times, against 9 to 12\% at the peak of a correction dose and none between doses. The beta cell argument therefore selects against the correction scale rather than for it. All figures here are sparing arithmetic from the model; we claim no clinical benefit for any of them.

\subsection*{Insulin sensitivity is not stationary}
Inverting the model for a per-day insulin sensitivity, holding secretory capacity fixed and using the charted dextrose, requires $\ISF$ of 67, 47, 91, 93, 61, 91 and 104~mg/dL per unit on successive days from day +42 to day +48: a rise of about half across the week, with two acute dips of about a third. Mean glucose fell 42~mg/dL from the first to the last of these days, of which the dextrose reduction accounts for 19 and the changing plant for 23. A fixed correction factor can track neither; a loop with integral action can track both.

\section*{Discussion}
The findings point the same way. The gain must not be fixed, the loop cannot act on individual readings, and only a term that does not depend on the measured state can move the operating point. We therefore suggest:
\begin{enumerate}
\item a continuous insulin infusion co-delivered with the dextrose at rate $u_b$, so that it stops automatically when the dextrose stops, removing the exogenous contribution to hypoglycemia during infusion interruptions;
\item no correction boluses, since a bolus produces a transient after which the patient returns to the operating point set by the carbohydrate load, and since a four-hour check cannot distinguish a sustained rise from a one-hour vehicle pulse;
\item a daily update
\begin{equation}
u_b \leftarrow u_b + \kappa\,\frac{\bar G_{\mathrm{prev}} - G_{\mathrm{target}}}{|dG/du|},
\end{equation}
where $\bar G_{\mathrm{prev}}$ is the previous day's mean of the routine checks, $G_{\mathrm{target}}$ the desired mean, $|dG/du|$ the sensitivity from Eq.~(\ref{eq:sens}) and $\kappa \approx 0.4$ a step-size gain, with a limit on the change per day. The daily mean is the one statistic the measurement can estimate;
\item rescue only below 70~mg/dL, treated with dextrose.
\end{enumerate}
This is integral control operating at a bandwidth below the measurement bandwidth, which is what a four-hour proportional rule gets backwards. It sits alongside rather than within the closed-loop literature\cite{Steil2011,Boughton2019}: those systems close the loop at minutes on a continuous sensor, whereas the question here is what can be done with four-hour sampling in a patient whose own loop is intact. Insulin added to the parenteral nutrition bag and separate continuous infusions are established practice\cite{Schonenberger2022}, but we know of no published protocol that updates a basal rate once daily on the previous day's mean, and the scheme is untested.

In simulation at 8~g/kg/day of dextrose with unbridged nutrition interruptions, the infusion brought the last three days' mean to 144~mg/dL against a target of 140, while the scale left it at 154 (Table~\ref{tab:sim}). With a deterministic plant, the same infusion run on a separate line produced glucose below 70~mg/dL in 28\% of runs during interruptions, and co-delivered in none. With the fitted sensitivity disturbance, readings below 70 occurred in 96 to 100\% of runs in every arm, including with no insulin at all; that disturbance overstates downward variability (Methods), and in that setting co-delivery cannot remove lows driven by the patient's own 37 units a day. Only protecting the dextrose supply can.

A further consideration bears on whether any correction is warranted. Inpatient targets are 140 to 180~mg/dL because tighter control has not shown benefit, and NICE-SUGAR found that targeting 81 to 108~mg/dL increased mortality and raised severe hypoglycemia fourteen-fold\cite{Finfer2009}. The index patient's checks at the current dextrose averaged 145, and no outcome benefit has been demonstrated for treating mild hyperglycemia in a non-diabetic adolescent on corticosteroids, so the scale was being used to push her below a target she already met. For the index patient the corollary is simpler still: the ordered scale is worth under 1~mg/dL and delivers under a unit a day, and its doses are triggered by readings that cannot distinguish a sustained rise from a vehicle pulse. Moving that vehicle to saline, or lengthening its infusion, would do more for her glucose than any change to the scale.

\paragraph{Limitations.} The analysis is a model calibrated to one patient and generalized by simulation; every quantitative figure is model output. Her true insulin sensitivity was never measured, and from level data alone it is identifiable only jointly with secretory capacity, which is why we report a per-day sensitivity trajectory rather than a constant. The size of the potency mismatch depends mainly on the secretion midpoint $G_{50}$, which was not measured in this patient; we used 170~mg/dL, above the 142~mg/dL of perifused human islets\cite{Buchwald2019}, and the virtual population shows that at islet-like values the mismatch is about 30\%, about half of it from the renal term. The renal term is the least certain entry: a urinalysis showed urine glucose above 1000~mg/dL, which at her measured output of 3.6 to 4.9~L/day implies at least 21 to 28~mg/dL/h of urinary loss, while the model predicts none below the textbook threshold of 180~mg/dL. If that loss is real, then at the same glucose the insulin-mediated term falls to 124 to 130~mg/dL/h and endogenous secretion to 31 to 33 units a day; we have not reconciled the two, and report potency with and without the renal term throughout. Glucose effectiveness is taken at the midpoint of its measured range\cite{Bergman1979}; at the extremes, 0.020 and 0.030 per minute, $\lambda$ is 4.9 and 5.5~$\mathrm{h}^{-1}$ and no conclusion turns on the choice, and a larger distribution volume or a two-compartment representation\cite{Caumo1998,Cobelli1999} acts in the same direction. The stochastic disturbance fitted to the within-day spread of the checks produces more low readings than were observed (simulated lowest check over nine days, median 76~mg/dL, against 110 observed), so hypoglycemia rates under that disturbance are upper bounds; the deterministic results bracket them from below. The correlation analysis rests on 45 observations, and the 12-hour result on 15; these are exploratory. Point-of-care error is taken as a standard deviation of 7~mg/dL. The vehicle-pulse excursions themselves have not been measured. Finally, this patient receives intravenous immunoglobulin, and maltose-containing preparations produce falsely elevated readings on glucose dehydrogenase pyrroloquinoline quinone strips, with errors large enough to have caused deaths\cite{Cohen2010}; we have not confirmed the strip chemistry in use, and if it is that chemistry the entire record is biased upward. The same interference is a competing explanation for the urinary glucose discrepancy, alongside drug-induced proximal tubular injury\cite{Herlitz2010}. A C-peptide with a simultaneous glucose, which would measure secretion directly and resolve the identifiability limitation, was requested and declined on the grounds that her catabolic state would make the result uninterpretable.

\section*{Methods}

\subsection*{Glucose model}
A single glucose compartment of volume $V_d = 0.16\,W$ liters, with every flux derived from the orders or from published physiology rather than fitted:
\begin{equation}
\begin{split}
\frac{dG}{dt} = {}& R_a(t) + \HGP(I, G) - U_{ni}(G) - k_{\mathrm{ren}}(G - G_{\mathrm{ren}})^+\\
& - S_G(G - G_{\mathrm{op}}) - \ISF(t)\, I, \qquad I = I_e + B(t),
\end{split}
\label{eq:model}
\end{equation}
\begin{equation}
\begin{split}
&\frac{dI_e}{dt} = \frac{S(G) - I_e}{\tau_e}, \qquad S(G) = \frac{s_{\max}}{1 + e^{-(G - G_{50})/\sigma_\beta}},\\
&\HGP(I, G) = \frac{\HGP^{\max}}{1 + (I/K_h)^2}\left(1 + \frac{(80 - G)^+}{30}\right).
\end{split}
\label{eq:secretion}
\end{equation}
Glucose rises from the infused dextrose $R_a$ and from the liver, and falls by brain and red cell consumption $U_{ni}$, by spillage into urine above the renal threshold, by glucose effectiveness and by insulin-driven uptake. $S(G)$ is the pancreas, a saturating sigmoid, and $\tau_e$ the delay between secretion and effect in tissue. Hepatic output is suppressed by insulin with a Hill exponent of 2\cite{Hill1910,Rizza1981}; suppression of hepatic glucose production is half-maximal at insulin levels well below those needed to stimulate peripheral uptake, and the dose-response is sigmoidal rather than hyperbolic\cite{Rizza1981}. The bracketed term is the counter-regulatory rise below 80~mg/dL; measured glycemic thresholds for glucagon and adrenaline release lie near 65 to 70~mg/dL\cite{Mitrakou1991}, so 80 is an early and conservative onset. $B(t)$ is the action of injected subcutaneous insulin, the impulse response of two first-order lags in series, $H(s) = [(1 + s\tau_{\mathrm{on}})(1 + s\tau_{\mathrm{off}})]^{-1}$. The loop closes through $I$: glucose sets secretion through $S(G)$, secretion becomes $I_e$ after the lag, and $I_e$ re-enters Eq.~(\ref{eq:model}) as insulin-mediated disposal. Parameters are listed in Table~\ref{tab:params}.

\begin{table}[h]
\centering
\caption{\textbf{Model parameters for the index patient.} $\ISF$ and $s_{\max}$ are not separately identifiable; $\ISF$ is set within the published glucocorticoid range and $s_{\max}$ to reproduce the operating point.}
\label{tab:params}
\small
\begin{tabular}{lp{5.0cm}p{5.6cm}}
\toprule
Symbol & Meaning & Value or source\\
\midrule
$W$ & body weight & 45.4 kg (dosing weight 48.1)\\
$V_d$ & glucose distribution volume & $0.16\,W = 7.26$ L\\
$R_a$ & glucose appearance from nutrition & dextrose order\\
$U_{ni}$ & insulin-independent uptake, brain and red cells & 1.4 mg/kg/min, Michaelis constant 20 mg/dL\cite{Baron1988,Gottesman1983}\\
$S_G$ & glucose effectiveness & 1.5 h$^{-1}$ (0.025 min$^{-1}$)\cite{Bergman1979}\\
$\HGP^{\max}$ & maximal hepatic output & 2.5 mg/kg/min\\
$k_{\mathrm{ren}}$, $G_{\mathrm{ren}}$ & renal clearance, threshold & $0.75\,\mathrm{GFR}/V_d$, GFR 113 mL/min from cystatin C; 180 mg/dL\\
$s_{\max}$ & maximal secretion rate & 5.20 U/h, set to give 139 mg/dL at 6 g/kg/day\\
$G_{50}$, $\sigma_\beta$ & midpoint and width of $S(G)$ & 170, 35 mg/dL\\
$\tau_e$ & secretion-to-effect lag & 0.4 h\\
$\tau_{\mathrm{on}}$, $\tau_{\mathrm{off}}$ & subcutaneous insulin lags & 0.6, 2.2 h\\
$K_h$ & insulin level halving hepatic output & $0.6\,\mathrm{U/h} \times \ISF_{\mathrm{norm}}/\ISF$\\
$\ISF$ & insulin sensitivity factor & 95 mg/dL/U, 37\% below $\ISF_{\mathrm{norm}} = 150$\\
$\CF$, $T$ & correction factor, target & 40 mg/dL/U; 125 day, 175 night\\
\bottomrule
\end{tabular}
\end{table}

\paragraph{Insulin-independent disposal.} Two terms carry it and they do different work. $U_{ni}$ is the flux taken by brain and red cells at 1.4~mg/kg/min\cite{Baron1988,Gottesman1983}; non-insulin-mediated uptake is roughly 80\% of whole-body glucose disposal in the fasting state. Those transporters are saturated at physiological glucose, which is why we model them with a Michaelis term and why their slope is only 0.04~$\mathrm{h}^{-1}$. The remainder of insulin-independent disposal is not saturated: hepatic uptake through glucokinase, basal transport in muscle and fat, and splanchnic and renal metabolism all rise roughly in proportion to glucose. Those routes are measured together as glucose effectiveness $S_G$, the rate at which glucose promotes its own disposal and suppresses its own production with insulin held at basal, obtained from the minimal model\cite{Bergman1979}. Lean normal subjects give 0.020 to 0.030 per minute; we take the midpoint. Minimal-model estimates are biased upward\cite{Caumo1998,Cobelli1999}, so the true value is more likely below the midpoint than above, which would raise the loop gain. Because measured $S_G$ already includes brain and red cell uptake, adding $U_{ni}'$ counts that contribution twice, 0.04 of 1.54 or under 3\%; we keep both so that the flux budget reads cleanly.

\paragraph{The operating point.} $G_{\mathrm{op}}$ is the glucose at which the fluxes of Eq.~(\ref{eq:model}) cancel: 139~mg/dL at 6~g/kg/day of dextrose, her steroid dose and her assumed sensitivity. It is not a property of the patient. It moves to 154~mg/dL at 8~g/kg/day and to 119 if her sensitivity returns to normal. The glucose-effectiveness term is referenced to it, so it counts only the change in insulin-independent clearance as she moves off the operating point and leaves the flux budget untouched.

\paragraph{Remaining parameters.} $R_a$ is computed from the prescribed dextrose on the dosing weight. $k_{\mathrm{ren}}$ uses her cystatin C based filtration rate. The distribution volume of 0.16~L/kg is at the low end of published tracer and volume-kinetic estimates, which span roughly 0.16 to 0.25~L/kg. Hepatic insulin resistance is tied to peripheral resistance through $K_h = K_h^{\mathrm{norm}} \ISF_{\mathrm{norm}}/\ISF$, with $\ISF_{\mathrm{norm}} = 150$~mg/dL/U a nominal value for an insulin-sensitive adolescent; at the chosen $\ISF$ of 95 that ratio is 1.6. Measured glucocorticoid-induced reductions in insulin action span roughly 30 to 60\%\cite{Rizza1982,Radhakutty2018}, and methylprednisolone at 2~mg/kg/day given twice daily is a high and sustained exposure, so the true reduction may exceed the 37\% assumed.

\subsection*{Patient data and calibration}
The index patient is a 17-year-old female, 45.4~kg, 41 to 51 days after allogeneic hematopoietic stem cell transplantation, with grade III gastrointestinal graft-versus-host disease, nothing by mouth, receiving all nutrition by continuous total parenteral nutrition and treated with methylprednisolone 2~mg/kg/day. Her glycated hemoglobin before transplant was 4.9\%. The record comprises 45 point-of-care glucose values from day +41 to day +51, the nutrition, medication and insulin orders, and the charted doses. Dextrose was 8~g/kg/day to day +44, 7 on days +45 and +46 and 6 from day +47. $\ISF$ and $s_{\max}$ enter the steady state only jointly. We set $\ISF$ to 95~mg/dL per unit, a 37\% reduction from the nominal 150 and within the 30 to 60\% reported for glucocorticoids\cite{Rizza1982,Radhakutty2018}, and $s_{\max}$ so that the model's steady state at 6~g/kg/day is 139~mg/dL; the per-day inversion below gives 91 and 104~mg/dL per unit on the two full days at that dextrose. Every other parameter comes from an order, a measured value or the literature (Table~\ref{tab:params}). The per-day sensitivity trajectory was obtained by solving, for each day with at least four checks spanning at least 11 hours, for the $\ISF$ at which the model's steady state equals that day's mean at that day's dextrose.

\subsection*{Numerical integration and potency}
Equations~(\ref{eq:model}) and (\ref{eq:secretion}) were integrated by explicit Euler with a step of 0.002~h, with secretion updated from the post-step glucose. Peak-response quantities converge slowly in the step: at 0.05~h $k(110)$ and $k(190)$ read 10.8 and 5.4~mg/dL per unit, at 0.002~h 11.6 and 5.8, and at 0.0005~h 11.7 and 5.8, while their ratio is 2.01 throughout. $k(G)$ was computed by holding the model at steady state $G$ with the dextrose that balances Eq.~(\ref{eq:model}) there, obtained in closed form, giving one unit subcutaneously, and recording the largest fall over eight hours.

\subsection*{Sampled-loop simulation}
Stochastic simulations used the same equations with a step of 0.005~h, at which $k(G)$ is within 0.7\% of its converged value. Insulin sensitivity was multiplied by $e^{x(t)}$, where $x$ is an Ornstein-Uhlenbeck process with a 6-hour correlation time. Its amplitude was fitted by bisection so that the pooled within-day standard deviation of simulated checks equals that of the 45 observed checks (22.8~mg/dL), giving 0.88; results are reported with and without it. Checks were taken near 00, 04, 08, 12, 16 and 20~h with uniform $\pm 0.5$~h jitter and meter error of standard deviation 7~mg/dL, within the 5 to 8~mg/dL implied by ISO 15197 and typical hospital meter coefficients of variation. Doses followed Eq.~(\ref{eq:rule}) with the charted targets. Vehicle infusions of 13.35~g at 06:00 and 8.35~g at 18:00, each over one hour, were included. Nutrition interruptions, where modeled, arrived as a Poisson process with a mean spacing of 48 hours and lasted 1 to 2 hours with no dextrose. The proposed infusion was updated at 06:00 each day with $\kappa = 0.4$, a limit of 0.15~U/h per day and a target of 140~mg/dL; co-delivered, it stopped with the dextrose. Each condition used 300 runs of ten days (fourteen for the interruption comparison), the first day discarded.

\subsection*{Virtual population}
We sampled 5,000 parameter sets uniformly from: body weight 40 to 90~kg; distribution volume 0.16 to 0.25~L/kg; glucose effectiveness 1.2 to 1.8~$\mathrm{h}^{-1}$\cite{Bergman1979}; insulin-independent uptake 1.4 to 1.6~mg/kg/min\cite{Baron1988,Gottesman1983}; maximal hepatic output 2.0 to 3.0~mg/kg/min; secretion midpoint $G_{50}$ 120 to 200~mg/dL; secretion width $\sigma_\beta$ 25 to 45~mg/dL; secretion lag 0.25 to 0.6~h; $\tau_{\mathrm{on}}$ 0.4 to 0.8~h and $\tau_{\mathrm{off}}$ 1.5 to 3.0~h; glucocorticoid reduction in insulin action 30 to 60\%\cite{Rizza1982,Radhakutty2018}; filtration rate 80 to 130~mL/min with renal scaling 0.5 to 1.0; and uncorrected operating point 130 to 200~mg/dL. Endogenous secretion at the operating point was drawn from $0.9 \pm 0.2$~U/kg/day\cite{Kruszynska1987,Polonsky1988}, truncated to 0.5 to 1.5. $\ISF$ was scaled inversely with distribution volume from the nominal 150~mg/dL/U and reduced by the glucocorticoid factor, $K_h$ scaled with weight and resistance, and $s_{\max}$ set so that secretion at the operating point equals the drawn value. The dextrose load that makes the operating point a steady state then follows in closed form; five sets implying less than 1 or more than 12~g/kg/day were excluded. For each patient we computed $\lambda$, the pancreatic share $A S'/\lambda$ and $dG/du$ at the operating point from Eq.~(\ref{eq:sens}), and $k(G)$ at 110 to 210~mg/dL by simulation, with and without the renal term, integrating all patients together at a step of 0.002~h. The implementation reproduces the index patient's $k(G)$ from the single-patient code to within 0.01~mg/dL per unit. Parameter influence was measured by partial rank correlation.

\subsection*{Correlation analysis}
Within-day residuals were formed by subtracting each day's mean. The lag-one correlation used consecutive pairs 3 to 5 hours apart. The slotted correlogram averaged products of residual pairs whose true separation fell in each bin, normalized by the residual variance, keeping bins with at least six pairs, with bands of $\pm 2/\sqrt{n}$. Folding replaced each separation by its distance to the nearest multiple of 24 or 12 hours. The 24-hour fold used the within-day residuals; the 12-hour fold used deviations from the mean of the window analyzed, since removing daily means from four to six checks a day would remove much of a 12-hour rhythm.

\subsection*{Ethics and consent}
Data were obtained from the existing clinical record of a single patient and analyzed retrospectively. No research intervention or additional data collection was performed, and formal ethics committee review was not sought. Written informed consent for publication was obtained from the patient's parent(s), and written assent was obtained from the patient. Direct identifiers, calendar dates and the name of the treating institution have been removed. The study was conducted in accordance with the Declaration of Helsinki.

\subsection*{Statistics and reproducibility}
The patient data comprise 45 point-of-care glucose measurements over eleven days. Correlation bands are $\pm 2$ standard errors; the 12-hour folding result uses the 15 checks at the current dextrose and is exploratory. Simulation results are medians or means over 300 runs per condition with fixed random seeds, and the virtual population uses a fixed seed; seeds are set in the code. Insulin sensitivity was set within the published glucocorticoid range and secretory capacity to reproduce the operating point; only the disturbance amplitude was fitted to data, and all other quantities come from orders or the literature.

\section*{Data availability}
The de-identified glucose series (post-transplant day and clock time), the dextrose schedule, the dosing record and the virtual population (one row per patient with all parameters and outputs) are available from the corresponding author on request, and will be deposited with the code on publication. Source data for Figs.~\ref{fig:potency}, \ref{fig:vpop} and \ref{fig:record} are generated by the figure script from these files.

\section*{Code availability}
The model, analysis and figure code (Python 3 with NumPy, SciPy, pandas and Matplotlib) are available from the corresponding author on request and will be deposited in a public repository on publication. Each numbered analysis script reproduces one set of results; \texttt{09\_virtual\_population.py} generates the virtual population and \texttt{10\_sampled\_loop.py} the stochastic simulations.

\section*{Acknowledgements}
This work received no external funding.

\section*{Author contributions}
N.R. conceived the study, developed the model, performed the analysis and wrote the manuscript. G.R. contributed to model development and reviewed and edited the manuscript. Both authors approved the final version.

\section*{Competing interests}
N.R. and G.R. are the owners of Regev Labs, LLC, which holds a provisional US patent application (64/160,853) on patient-specific glucose plant models and insulin correction rules, and a second provisional application on insulin infusion scheduling. N.R. and G.R. are family members of the patient. The authors declare no other competing interests.


\end{document}